# Storage and Caching: Synthesis of Flow-based Microfluidic Biochips

Tsun-Ming Tseng[1]   Bing Li[1]   Tsung-Yi Ho[2]   Ulf Schlichtmann[1]

[1]Technische Universität München
Arcisstrasse 21, 80333 Munich, Germany

[2]National Tsing Hua University
No. 101, Section 2, Kuang-Fu Rd., Hsinchu, Taiwan

## ABSTRACT

Flow-based microfluidic biochips are widely used in lab-on-a-chip experiments. In these chips, devices such as mixers and detectors connected by micro-channels execute specific operations. Intermediate fluid samples are saved in storage temporarily until target devices become available. However, if the storage unit does not have enough capacity, fluid samples must wait in devices, reducing their efficiency and thus increasing the overall execution time. Consequently, storage and caching of fluid samples in such microfluidic chips must be considered during synthesis to balance execution efficiency and chip area.

## 1. INTRODUCTION

Microfluidic biochips have revolutionized traditional biochemical diagnoses and chemical experiments significantly by enabling lab-on-chips. On such a chip, samples and reagents are propagated and mixed in volumes of nanoliters instead of large drops. This miniaturization significantly saves reagents, which are very expensive in many experiments. In addition, the experimental flow on such a chip is controlled by a microcontroller, so that the schedule of each operation in the assay is maintained accurately. Consequently, both the execution time of the assay and the quality of the experiment are improved.

Flow-based microfluidic biochips have dedicated devices such as mixers and detectors for specific operations. These devices are connected by micro-channels, through which fluid samples and reagents are transported from one device to another.

Micro-channels are made from dimethylsiloxane using soft lithography. The transportation of fluid samples through these channels is controlled by valves, whose basic structure is shown in Figure 1a. In such a structure, a flow channel is constructed on a substrate to transport fluid samples and reagents. Above the flow channel, a control channel is constructed and connected to an air pump. Since both channels are built from elastic materials, an air pressure applied in the control channel squeezes the flow channel tightly, so that the movement of the fluid sample can be blocked. Reversely, if the pressure in the control channel is released, the fluid sample can resume its movement to the target device.

Valves can be used to construct more complex devices. In biochips, it is very common that transportation routes of several fluid samples cross each other. At such a crossing point, a switch can be constructed using valves, as shown in Figure 1b. In this device, only a pair of valves are open at the same time to direct the fluid sample. Another dedicated device is a mixer, as shown in Figure 1c. In this device, the three valves at the top are actuated alternately to create a circular flow around the device to mix different fluid samples.

After an operation is finished, the intermediate result can be transported to other devices or saved temporarily in a dedicated storage unit. Figure 1d shows a detailed schematic of a mixer connected to a storage unit with eight cells [1]. These side-by-side storage cells are constructed from normal flow channels but with multiplexer-like controlling valves at each end. Consequently, only one fluid sample can enter or leave the storage unit at a certain moment.

A biochip executes operations in an assay by time multiplexing. Such an assay is usually specified by a sequencing graph. In Figure 1e the sequencing graph of polymerase chain reaction (PCR) is shown. This assay takes eight input samples ($i_1 \sim i_8$) and mixes them with seven operations ($o_1 \sim o_7$) to generate copies of DNA sequence. If for each operation a mixer is assigned, seven mixers should be built on the chip. However, to reduce cost it is not usual to assign resource so freely. Instead, mixers are reused to execute the operations while maintaining their dependency specified by the sequencing graph. For example, a mixer can be used repeatedly to execute the operations in Figure 1e. Meanwhile, intermediate reaction samples such as the output of $o_1$ can be saved in a storage unit until the result of $o_2$ is available. With this time multiplexing, the number of mixers on the chip can be reduced significantly.

To execute an application on a flow-based biochip efficiently, the operations in the application should be carefully assigned to specific devices in the chip in proper time slots, thus requiring a complete flow of design automation.

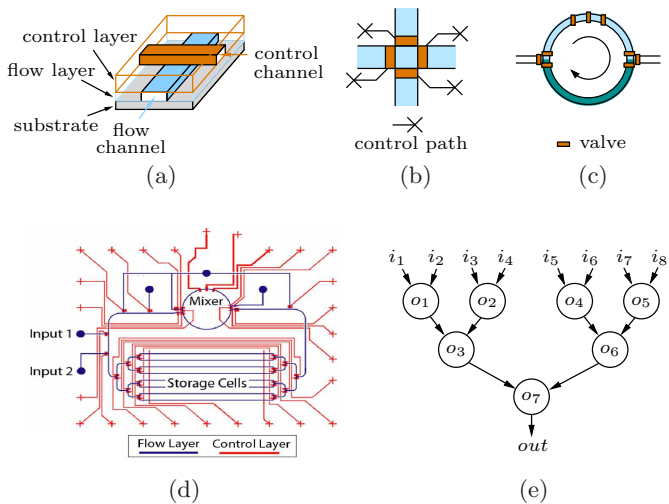

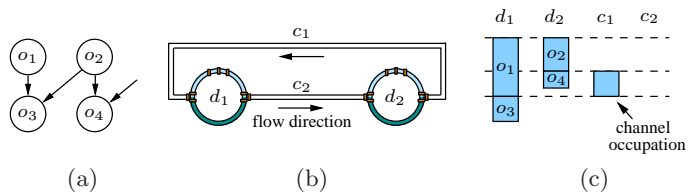

Figure 1: (a) Valve structure. (b) Switch. (c) Mixer. (d) A biochip with a mixer and an eight-cell storage unit, adapted from [1]. (e) Sequencing graph of the PCR assay.

Figure 2: Channel caching. (a) Sequencing graph. (b) Mixer and channel structure. (c) Schedule with caching.

Owing to advances of fabrication techniques, the integration of flow-based biochips is continuously increasing [2]. Consequently, research in this area has attracted much attention. In [3] a top-down flow for architectural synthesis is proposed. In [4] control layer synthesis is addressed. In addition, fault modeling and test generation are covered in [5], and dynamic mapping on a valve array is discussed in [6].

In this article, we introduce a new concept to use transportation channels connecting dedicated devices as temporary caches of fluid samples. Sample assignment in dedicated and distributed storage is also discussed. The goal of this synthesis process is to reduce both the overall execution time of the assay and the chip area at the same time. This is the first work considering channel caching in flow-based biochips, and new constraints are introduced to avoid channel conflicts.

## 2. STORAGE AND CACHING

In a traditional biochip, there is usually one dedicated storage unit and it should be considered directly for an optimized design. Consider Figure 1e which depicts the mixing stage of the PCR assay and will be executed by the biochip in Figure 1d. Assume that all the operations in this assay have the same execution time. During the execution, the resulting fluid sample from an operation needs to be saved in a storage unit if its child operation is not the next one to be executed. For example, after the execution of $o_1$ and $o_2$, both output samples should be saved in the storage unit if the next operation to be executed is $o_4$. But if the next operation is $o_3$, the output of $o_2$ can stay in the mixer directly, while the result of $o_1$ is fetched from the storage unit. Consequently, not only transportation time but also storage usage are reduced.

In the example above, there are only one mixer and one storage unit. For large biochemical assays, more mixers are integrated into the chip to reduce the execution time. This increased parallelism produces more concurrent intermediate samples that should be saved in the storage unit. In traditional design methods, this is achieved by increasing the capacity of the storage unit, namely the number of storage cells. Consequently, a large monolithic storage unit containing many cells is formed. Since a storage unit has only one input port and one output port, and allows only one fluid sample to enter or leave, the competition for these ports by multiple samples may hurt the performance of the chip in the end, largely negating the benefit of increasing the number of dedicated devices.

In addition to the dedicated storage unit, transportation channels themselves can also be considered as temporary caching cells to reduce the execution time. An example of this caching usage is illustrated in Figure 2. In this example, operations $o_1$ and $o_2$ are executed at the same time in mixers $d_1$ and $d_2$, respectively. As $o_2$ finishes earlier than $o_1$, device $d_2$ can dump the resulting fluid sample of $o_2$ into channel $c_1$ and then start the operation $o_4$. In existing methods, channels are only used to transport fluid samples and not considered as temporary caching cells. Therefore, the result of $o_2$ can be transported to $d_1$ only after $o_1$ is finished and $o_4$ can start only then.

## 3. SYNTHESIS WITH CHANNEL CACHING AND STORAGE ASSIGNMENT

In this section, constraints for scheduling and binding of a biochemical assay are reviewed briefly. Thereafter, additional constraints to avoid fluid conflicts when applying channel caching are introduced. These constraints together with the basic scheduling and binding constraints are solved as a whole using an ILP solver. In the last step, intermediate fluid samples that cannot be handled by channels are assigned to an external storage unit by time-multiplexing. The last step is a post-processing step, dealing with the result produced by the ILP solver. This step offloads some constraints from the ILP formulation

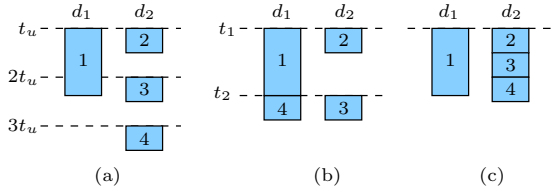

Figure 3: Model comparison. (a) Stage-based model [7,8]. (b) Stage-based model with variable-length time slots [9]. (c) Stageless model.

above to reduce the problem complexity.

## 3.1 Common constraints

Since the number of devices in a biochip is usually smaller than the number of operations in an assay, more than one operation is assigned to a device while maintaining the dependency specified by the sequencing graph, i.e., a child operation should be executed later than its parents. Therefore, the synthesized result must meet the following constraints [3, 7, 8].

**Uniqueness** : An operation should be scheduled to a device only once.

**Duration** : The difference between the starting time and the ending time of an operation should be no smaller than its duration.

**Dependence** : A child operation should not start before the results from its parents arrive.

**Non-overlapping operations** : Any two operations that are overlapping along the timeline should not be assigned to the same device.

The non-overlapping constraints are created by comparing the starting and ending times of each pair of operations. This is different from the stage-based scheduling models in [7–9] as demonstrated in Figure 3, where four operations are assigned to two devices. The classical stage-based model in Figure 3a schedules operations to pre-determined time slots [7,8]. This model is revised in [9] to allow variable-length time slots for pin-count reduction, as illustrated in Figure 3b. In the proposed method, the direct comparison of starting and ending times of operations is actually a stageless model shown in Figure 3c, which allows an operation to start at any time so that operations are packed tightly to reduce the overall execution time of the assay.

## 3.2 Channel caching

In existing scheduling and binding methods for flow-based biochips, the transportation time between devices is not modeled directly. After a device finishes an operation, the result is supposed to be saved in storage and fetched back when its child operation is ready. This simplification may lead to a significant performance drop because fluid samples might have many conflicts at the ports of the storage unit.

Using channels to cache intermediate samples can alleviate the port competition problem above significantly. This usage requires to include *channel conflict constraints* into scheduling and binding. Since a channel can only be used by one fluid sample, such a constraint requires that a new fluid sample should not be dumped into a channel which is still occupied by a previous sample. Otherwise, contamination between fluid samples occurs. For example, in Figure 2c operation $o_4$ executed by device $d_2$ finishes earlier than operation $o_1$. But device $d_2$ cannot dump its result into channel $c_1$ because this channel is still occupied by the output of operation $o_2$.

To avoid contamination in channels, we model the conflict scenarios as illustrated in Figure 4, where $o_{i_1}$ and $o_{i_2}$ are executed by device $d_{k_1}$, and $o_{j_1}$ and $o_{j_2}$ are executed by $d_{k_2}$. In the case on the left, the result of $o_{i_2}$ may contaminate the result of $o_{i_1}$ because the latter has not entered the device $d_{k_2}$ and still occupies the channel. Similarly, the case on the right shows the mirrored case where $o_{i_2}$ is executed earlier than $o_{i_1}$. Consequently, transportation requests of the two edges $(o_{i_1}, o_{j_1})$ and $(o_{i_2}, o_{j_2})$ compete for the channel between $d_{k_1}$ and $d_{k_2}$, and one of these requests should be directed to the dedicated storage unit. A 0-1 variable $\lambda_{i_1,k_1,j_1,k_2}$ is defined to represent the use of the dedicated storage unit when $o_{i_1}$ and $o_{j_1}$ are mapped to devices $d_{k_1}$ and $d_{k_2}$, respectively, as

$$\lambda_{i_1,k_1,j_1,k_2} = \begin{cases} 1 & \text{if the output of } o_{i_1} \text{ should be directed to} \\ & \text{the dedicated storage unit.} \\ 0 & \text{otherwise.} \end{cases} \quad (1)$$

To avoid channel conflicts, either the operation $o_{i_2}$ finishes later than $o_{j_1}$ starts or similarly $o_{i_1}$ finishes later than $o_{j_2}$ starts. Therefore, the non-conflict condition of channel usage can be expressed as

$$\forall \text{ edge } (o_{i_1}, o_{j_1}), \text{edge } (o_{i_2}, o_{j_2}), \text{device } d_{k_1}, \text{and device } d_{k_2}$$
$$\text{if } e_{i_2} < t_{j_1} \text{ and } e_{i_1} < t_{j_2}$$
$$\lambda_{i_1,k_1,j_1,k_2} + \lambda_{i_2,k_1,j_2,k_2} \geq 1 \quad (2)$$

where $t_{j_1}$ and $t_{j_2}$ are the scheduled starting times of operations $o_{j_1}$ and $o_{j_2}$, respectively; $e_{i_1}$ and $e_{i_2}$ are the scheduled ending times of operations $o_{i_1}$ and $o_{i_2}$, respectively. The condition $e_{i_2} < t_{j_1}$ and $e_{i_1} < t_{j_2}$ describes the two conflict cases in Figure 4. The sum in (2) describes that at least one output should be directed to the dedicated storage unit. This conditional constraint can be transformed into a linear form [10] and handled by an integer linear programming (ILP) solver.

In the proposed model, caching using channels does not incur any cost. Therefore, the more caching is conducted by channels, the smaller the dedicated storage unit becomes. According to this observation, the proposed method minimizes the number of conflicted sample transportations to be handled by the dedicated storage unit. This number is given by

$$\lambda_{ds} = \sum_{i_1,k_1,j_1,k_2} \lambda_{i_1,k_1,j_1,k_2}. \quad (3)$$

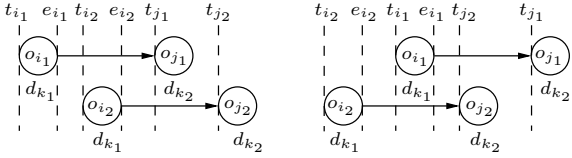

**Figure 4: Channel conflict scenarios.**

By minimizing $\lambda_{ds}$, fluid samples are forced to be cached in transportation channels as much as possible. Accordingly, the number of cells in the dedicated storage unit is decreased to reduce chip area.

### 3.3 Optimize channels and execution time

The concept of channel caching can be applied to a biochip with a given number of channels and their connections. In this case, the constraints (1)–(3) are created only for the available channels. If the number of channels between devices is not given, a full channel connection between any pair of devices are assumed. In both cases, the proposed method reduces the execution time of the application using as few channels as possible. The channels that are not used after synthesis are simply removed to save resources.

Assume that operations $o_i$ and $o_j$ have an edge in the sequencing graph, meaning that the result of $o_i$ should be transported to $o_j$. If these two operations are assigned to devices $d_{k_1}$ and $d_{k_2}$, there should be a channel between $d_{k_1}$ and $d_{k_2}$. We maintain a 0-1 variable $c_{k_1,k_2}$ to represent the presence of this channel. Consequently, the total number of channels in the chip can be constrained as

$$\sum_{d_{k_1},d_{k_2}} c_{k_1,k_2} \leq n_c \quad (4)$$

where $n_c$ is the upper bound of the number of channels.

Another major objective of the synthesis process is to minimize the execution time $T$ of the assay, which is determined by the latest finishing time of all the operations. Therefore, it is constrained as

$$\forall \text{ operation } o_i, \ e_i \leq T \quad (5)$$

In synthesizing a given assay to achieve a short execution time with as few channels as possible, the upper bound of the execution time $T$ in (5) and the upper bound of the number of channels $n_c$ in (4) should be minimized.

The overall optimization problem is summarized as follows:

$$\text{Minimize: } w_t T + w_c n_c + w_s \lambda_{ds} \quad (6)$$

Subject to:

$$\text{constraints described in Section 3.1 and} \quad (7)$$
$$\text{channel constraints (1)–(5)} \quad (8)$$

where the weight $w_t$ is set 1 and $w_c$ is set to 0.01, so that the execution time has a high priority to be reduced as much as possible. In practice, different pairs of weights may be used for a tradeoff.

### 3.4 Storage assignment

After the synthesis model above is solved, flow transportation requirements may still exceed the capacity of channel caching and some of them should be directed to the dedicated storage unit.

The period a fluid sample stays in the storage unit can be partitioned into three phases. In phase one, it enters the storage unit. Because a storage unit only allows one fluid sample to enter or leave due to the flow path, as illustrated in Figure 1d, only one fluid sample is allowed to be in phase one at a time. Otherwise, a *port conflict* occurs. In phase two, the fluid sample occupies a cell in the storage unit. In phase three, it leaves the storage unit, and again only one fluid sample is allowed to use the port of the storage unit. Figure 5a illustrates these three phases, and Figure 5b shows an example of four fluid samples directed to the dedicated storage unit but with port conflicts.

Since two fluid samples cannot enter the same storage unit if there is a conflict at phase one or phase three, the largest set of fluid samples that do not conflict with each other are identified and packed into the dedicated storage unit. The conflict relation can be represented using a graph as illustrated in Figure 5c, where nodes represent fluid samples. If there is a port conflict between any two fluid samples, an edge is created between the corresponding nodes. The problem is thus transformed to find the maximum independent set of nodes between which there is no edge. This problem is then solved using the algorithm in [11]. In this algorithm, the node with the smallest degree is selected and removed together with all the nodes connected to it. This process is repeated until all nodes are removed from the graph, and the selected nodes together form an independent set. For example, if node 2 together with node 1 in Figure 5c is removed, the independent fluid samples $\{s_2, s_3, s_4\}$ are identified as candidates to be saved in the dedicated storage unit.

Among the fluid samples directed to the storage unit, any pair of them that do not occupy storage cells at the same time can actually reuse the same cell in the style of time multiplexing. Similar to handling port conflicts, the maximum independent set of samples sharing the same storage cell can be found by the algorithm in [11]. In Figure 5d, the only edge represents that $s_2$ and $s_3$ have a conflict during phase two of storage. The result shows that $s_2$ and $s_4$ can share the same storage cell.

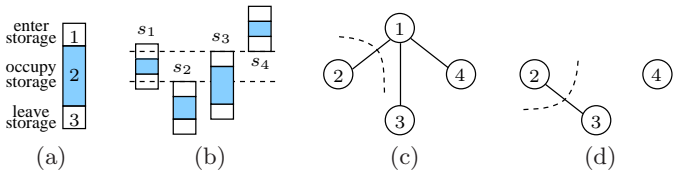

Figure 5: Storage assignment. (a) Fluid phases. (b) Fluid conflicts. (c) Port conflict graph. (d) Storage cell conflict graph.

Table 1: Results with storage and channel caching

| Assay | List alg. with storage | | | Storage and caching | | | |
|---|---|---|---|---|---|---|---|
| | $\#ch_L$ | $\#sto_L$ | $T_L$ | $\#ch_p$ | $\#sto_p$ | $T_p$ | $r_p$(s) |
| PCR7 | 1 | 2 | 49.73 | 1 | 1 | 49.73 | 0.16 |
| MT18 | 4 | 1 | 62.4 | 4 | 0 | 60.84 | 3.22 |
| PI39 | 6 | 7 | 129.94 | 5 | 5 | 98.57 | 17.14 |
| PE55 | 6 | 10 | 161.53 | 6 | 9 | 121.38 | 72.79 |
| gen15 | 4 | 4 | 41.64 | 3 | 1 | 39.73 | 15.73 |
| gen31 | 4 | 6 | 79.83 | 4 | 6 | 77.59 | 72.57 |
| gen63 | 16 | 19 | 83.39 | 14 | 11 | 81.62 | 440.35 |
| gen127 | 25 | 32 | 139.58 | 23 | 23 | 139.18 | 894.31 |
| average reduction | | | | 7.8% | 41.7% | 7.7% | |

After determining the independent sets, there might be a few fluid samples that cannot be assigned into the dedicated storage unit due to port conflicts. These samples are saved directly in additional distributed storage cells built along channels to generate the final chip structure.

## 4. RESULTS

The proposed method was implemented using C++, and tested on a computer with a 2.67 GHz CPU. Four real biochemical cases from [12] and four synthetic cases, gen15–gen127, were used for experiments. The List algorithm in [8], which does not consider constraints from channels and storage, was implemented for comparison. This algorithm produced schedules for the assays, to which the same maximum independent set algorithm in [11] was applied to generate distributed and dedicated storage units.

The experimental results are shown in Table 1. The columns $T_L$ and $T_p$ are the assay execution times calculated by the List algorithm and the proposed method. From this comparison, we can see that the proposed method resulted in improvements in almost all assays, by 7.7% on average. Specially for PI39 and PE55, the improvement on execution time can reach nearly 25%.

The results of transportation channels and storage cells are shown in columns $\#ch_L$ and $\#sto_L$, respectively. The results from the proposed method are shown in the columns $\#ch_p$ and $\#sto_p$, respectively. Obviously the proposed method does not require more channels or storage cells to achieve the shortened execution time. Actually in cases such as gen63, channels and storage cells are reduced significantly. On average, these reductions reach 7.8% and 41.7%, as shown in the last row of Table 1.

To demonstrate the effect of channel caching, a baseline method was implemented. In this method, a device cannot start a new operation before its previous output sample is taken by another device to avoid channel conflicts and thus sample contamination. The execution times calculated by the baseline method and the proposed method are illustrated in Figure 6. Clearly, the proposed method effectively reduces the execution time of an assay by simply caching fluid samples in transportation channels.

The runtimes of solving the proposed ILP model and the storage assignment for the test cases are shown in the column $r_p$(s) in Table 1. For the largest application with 127 operations, the runtime is 894.31 seconds, largely taken by the ILP solver. These computational runtimes are already acceptable for an offline synthesis flow.

## 5. CONCLUSION

In this paper, a concept to cache fluid samples in transportation channels and synthesize storage cells considering fluid conflicts is explained. By minimizing channel conflicts and recognizing maximum independent sets, storage requirements are handled jointly by channels as well as both distributed and dedicated storage cells. Results show that the execution time of the assay and resource usage are lowered effectively at the same time.

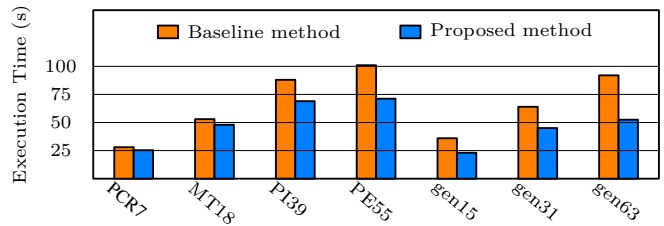

Figure 6: Execution times of assays calculated by the baseline method and the proposed method.